\documentclass[12pt,a4paper]{article}
\usepackage{epsfig}
\usepackage{amsmath}
\usepackage{amssymb}
\usepackage{verbatim}
\usepackage{bm}
\usepackage{dsfont}
\usepackage[footnotesize]{caption}
\usepackage{subfigure}
\usepackage{cancel}
\usepackage{cite}
\newcommand{\Mp}{M_{\mathrm{P}}}

\begin{document}

\begin{flushleft}
DESY 13-102\\
June 2013
\end{flushleft}

\vskip 1cm

\begin{center}
{\Large\bf The Starobinsky Model from\\[2mm]  Superconformal D-Term Inflation}

\vskip 2cm

{ W.~Buchmuller, V.~Domcke,  K.~Kamada}\\[3mm]
{\it{
Deutsches Elektronen-Synchrotron DESY, 22607 Hamburg, Germany}
}
\end{center}

\vskip 1cm

\begin{abstract}
\noindent 
We point out that in the large field regime, the recently proposed 
superconformal D-term inflation model coincides with the Starobinsky model. 
In this regime, the inflaton field dominates over the Planck mass in the 
gravitational kinetic term in the Jordan frame. Slow-roll inflation is
realized in the large field regime for sufficiently large gauge couplings. 
The Starobinsky model generally emerges as an effective description of
slow-roll inflation if a Jordan frame exists where, for large inflaton
field values, the action is scale invariant and the ratio $\hat{\lambda}$
of the inflaton self-coupling and the nonminimal coupling to gravity is 
tiny. The interpretation of this effective coupling is 
different in different models. In superconformal D-term inflation it is determined 
by the scale of grand unification, $\hat{\lambda} 
\sim (\Lambda_{\rm GUT}/\Mp)^4$. 
\end{abstract}.

\thispagestyle{empty}

\newpage

%\section{Introduction}

The recently released data from the Planck satellite provide a 
precise picture of the cosmic microwave background radiation 
\cite{Ade:2013uln}. The observed temperature anisotropies are consistent
with a primordial spectrum of density perturbations produced during an
inflationary phase \cite{Linde:2005ht}. In fact, the data support the 
simplest version of inflation, the single field slow-roll paradigm 
\cite{Ade:2013uln}. However, several popular
inflation models \cite{Martin:2013tda} are strongly disfavoured or even ruled 
out by the data. It is therefore remarkable that the first inflation model, 
the $R^2$ model of Starobinsky \cite{Starobinsky:1980te}, is fully consistent 
with the Planck data \cite{Ade:2013uln}.

Recently, a supergravity model of inflation has been proposed 
\cite{Buchmuller:2012ex}, which is based on the superpotential and
scalar potential of D-term hybrid inflation 
\cite{Binetruy:1996xj,Halyo:1996pp}, and a K\"ahler potential motivated
by the underlying superconformal symmetry of supergravity
\cite{Einhorn:2009bh,Ferrara:2010yw}. For such models,
there is a Jordan frame in which the matter part of the
Lagrangian takes a particularly simple form, closely resembling global
supersymmetry. Depending on gauge and Yukawa couplings, the model allows
for small field as well as large field inflation. In this note we point
out that in the large field regime the inflaton potential and the spectral
indices agree with the predictions of the Starobinsky model.

\vspace{0.6cm}
\noindent \textbf{Superconformal D-term inflation}
\vspace{0.3cm}

Let us briefly recall the main ingredients of the model proposed in
Ref.~\cite{Buchmuller:2012ex}.
Supersymmetric D-term hybrid inflation models contain two `waterfall' fields
$\phi_{\pm}$ and and an inflaton field $S$, with the superpotential
\begin{align}
W = \lambda S \phi_+ \phi_- \ .
\end{align}
In the superconformal version the K\"ahler potential\footnote{We use
units where $\Mp = 1/\sqrt{8\pi G} = 1$.} reads
($z^{\alpha}=\phi_{\pm},S$) 
\begin{align}\label{KahlerP}
K(z,\bar{z}) &= 3\ln \Omega^2(z,\bar{z}) \qquad \textrm{with} \nonumber\\
\Omega^{-2} &= 1 -\frac{1}{3}\left( |S|^2 + |\phi_-|^2 + |\phi_+|^2\right)  
- \frac{\chi}{6}\left(S^2 + \bar{S}^2\right) \ ,
\end{align}
where the holomorphic part proportional to $\chi$ breaks superconformal 
symmetry explicitly \cite{Einhorn:2009bh,Ferrara:2010yw}. 
In the Einstein frame with metric $g$, the Lagrangian reads
\begin{align}\label{einstein}
\frac{1}{\sqrt{-g}}\mathcal{L} = 
\frac{1}{2} R - K_{\alpha\bar{\alpha}}g^{\mu\nu}
\nabla_{\mu}z^{\alpha}\nabla_{\nu}\bar{z}^{\bar{\alpha}}  - V_D - V_F \ ;
\end{align} 
here $Q$ is the matrix of $U(1)$ charges, 
$\nabla_{\mu} = \partial_{\mu} -igA_{\mu}Q$ is the gauge covariant
derivative,
$K_{\alpha\bar{\alpha}} = \partial_{\alpha}\partial_{\bar{\alpha}}K$ and 
\begin{align}\label{DPotDterm}
V_D &= \frac{g^2}{2} \left(\partial_{\alpha}K Q z^{\alpha} 
+ \xi\right)^2 
= \frac{g^2}{2} \left(\Omega^2 q(|\phi_+|^2 -|\phi_-|^2)-\xi\right)^2 
\end{align}
is the D-term potential for charges $0$ and $\pm q$ of the chiral superfields 
$S$ and $\phi_\pm$, respectively; $\xi$ is the Fayet-Iliopoulos term, and 
the F-term scalar potential is given by
\begin{align}\label{PotEJ}
V_F &= \Omega^4 \left(
\delta^{\alpha\bar{\alpha}} \partial_{\alpha}W \partial_{\bar{\alpha}}\overline{W}\right) \nonumber\\
& = \Omega^4\lambda^2\left(|S|^2 (|\phi_+|^2 + |\phi_-|^2) 
+ |\phi_+ \phi_-|^2  -  \frac{\chi^2 |\phi_+|^2 |\phi_-|^2 |S|^2}
{3 + \frac{1}{2} \chi (S^2 + {\bar S}^2) + \chi^2 |S|^2}\right) \ .
\end{align}
On the inflationary trajectory one has $\phi_\pm = 0$. Hence $V_F$ vanishes 
identically and
$V_D$ provides the vacuum energy $V_{0} = g^2 \xi^2/2$ which drives inflation.

It is very instructive to also consider the theory in the Jordan frame
defined by the metric $g_{J\mu\nu} = \Omega^2 g_{\mu\nu}$. This 
Weyl transformation yields the Lagrangian \cite{Ferrara:2010yw}
\begin{align}\label{LagJ}
\frac{1}{\sqrt{-g_J}}\mathcal{L}_J &= 
\frac{1}{2}\Omega^{-2} R_J - \delta_{\alpha\bar{\alpha}}g_J^{\mu\nu}
\nabla_{\mu}z^{\alpha}\nabla_{\nu}\bar{z}^{\bar{\alpha}}  - V_J \ ,
 \nonumber\\
V_J &= \Omega^{-4}\left(V_D + V_F\right) \ .
\end{align} 
Contrary to the Einstein frame the kinetic term of the gravitational field
is now field dependent whereas the kinetic terms of the scalar fields are
canonical. Along the inflationary trajectory one has
\begin{align}\label{omega0}
\Omega_0^{-2} = \Omega^{-2}\big|_{\phi_\pm = 0} = 1 - \frac{1}{3}
\left(|S|^2 + \frac{\chi}{2}(S^2 + \bar{S}^2)\right) 
\ , \quad V_J = \Omega_0^{-4}\ \frac{g^2}{2} \xi^2 \ , 
\end{align}
i.e., the Fayet-Iliopoulos term is field dependent.

The slope of the inflaton potential is generated by quantum corrections.
A straightforward calculation yields for the one-loop potential, 
\begin{align}\label{eq_1-loop}
V_{1l} &= \frac{g^4 q^2\xi^2}{32\pi^2}\left((x-1)^2\ln(x-1) + (x+1)^2 \ln(x+1) 
- 2 x^2 \ln x  -1 \right) \nonumber \\
&= \frac{g^4 q^2\xi^2}{16\pi^2} \left(1 + \ln x 
+ {\cal O}\left(\frac{1}{x}\right) \right) \ , \quad 
x = \frac{\Omega_0^2(S) |S|^2}{\Omega_0^2(S_c) |S_c|^2}\ .
\end{align}
The critical field value $S_c$, where the mass of the waterfall
field $\phi_+$ reaches zero, is determined by
\begin{align}
 \Omega^2_0(S_c)|S_c|^2 = \frac{q g^2 \xi}{\lambda^2}\ .
\label{eq_Sc}
\end{align}
The total potential is then given by 
\begin{align}\label{hybridpotential}
V &= (V_F + V_D + V_{1l})\big|_{\phi_\pm = 0} \nonumber\\
&= \frac{g^2}{2}\xi^2 \left(1 + \frac{g^2 q^2\xi^2}{8\pi^2} 
\left(1 + \ln x + {\cal O}\left(\frac{1}{x}\right)\right)\right)\ .
\end{align}
Note that on the inflationary trajectory one has $|S| > |S_c|$ and $x > 1$.

\vspace{0.6cm}
\noindent \textbf{Single field slow-roll inflation}
\vspace{0.3cm}

Expressing the Lagrangian for the field $S$ in terms of real and imaginary
components, $S = (\sigma + i \tau)/\sqrt{2}$,
\begin{equation}
\frac{1}{\sqrt{-g}}\mathcal{L} = \frac{1}{2} K_{S \bar{S}}(\sigma,\tau) 
(\partial_{\mu} \sigma \partial^{\mu} \sigma + 
\partial_{\mu} \tau \partial^{\mu} \tau) - V(\sigma, \tau) \ ,
\label{eq_lagrange}
\end{equation}
one obtains the slow-roll equations for the homogeneous fields $\sigma$ 
and $\tau$,
\begin{equation}
 3 K_{S \bar{S}} H \dot \sigma = - \frac{dV_{1l}}{d \sigma}\ ,  \quad   
3 K_{S \bar{S}} H \dot \tau = - \frac{dV_{1l}}{d \tau}\ .
\label{eq_double_sr}
\end{equation}
One easily verifies that for $\chi < 0$, which we choose w.l.o.g., the 
trajectory $\sigma \neq 0, \tau = 0$ is an attractor for a sufficiently 
long phase of inflation before the onset of the final $N_*$ e-folds. 
Inserting the K\"ahler metric,
\begin{align}
K_{S \bar{S}}\big|_{\phi_\pm,\tau =0} = 
\frac{1}{1-\frac{1}{6}(1+\chi)\sigma^2}\left(1 +
\frac{(1+\chi)^2\sigma^2}{6\left(1-\frac{1}{6}
(1+\chi)\sigma^2\right)}\right) \ ,
\end{align}  
and the one-loop potential (\ref{eq_1-loop}) into the slow-roll equation
(\ref{eq_double_sr}), one obtains after integrating from  $\sigma_*$ to 
$\sigma_f$, 
\begin{equation}
 3 \ln \left( \frac{1 - \frac{1}{6}(1 + \chi) \sigma_*^2}{1 - \frac{1}{6}(1 + \chi) \sigma_f^2}\right) - \frac{1}{2} \chi \left(- \sigma_*^2 + \sigma_f^2\right) \simeq - \frac{g^2 q^2}{4 \pi^2} N_*  \ .
\label{eq_sigmae}
\end{equation}
Here $\sigma_f$ denotes the value of $\sigma$ at the end of inflation and 
$\sigma_*$ is the value of $\sigma$ $N_*$ e-folds earlier. 
Inflation ends when either $m^2_+$ turns negative,
\begin{align}
\sigma_f^2 = \sigma_c^2 
= \frac{6 g^2 q \xi}{3 \lambda^2 + (1 + \chi) g^2 q \xi}\ ,  
\label{eq_sigma_f}
\end{align}
or when the slow-roll conditions are violated, i.e. $\sigma_f = \sigma_{\eta}$,
for sufficiently large values of $\lambda$. 

For small couplings, $gq \ll 1$, inflation takes place at small field
values. In this case $-(1+\chi)\sigma_*^2/6 < 1$, and 
Eq.~(\ref{eq_sigmae}) implies
\begin{align}\label{smallfield}
\sigma_f^2 < \sigma_*^2 \simeq  \frac{g^2 q^2}{2 \pi^2} N_* < 1 \ .
%\label{eq_sigmae_simple}
\end{align}
Here we are particularly interested in the large field regime, 
$-(1+\chi)\sigma_*^2/6 > 1$, which is realized for large couplings $gq$.
As we shall see, for couplings in the perturbative regime, one 
typically has $-(1+\chi)\sigma_f^2/6 < 1$. From Eq.~(\ref{eq_sigmae}) 
one then obtains 
\begin{align}\label{largefield}
\sigma_f^2 < -\chi\sigma_*^2 \simeq  
\frac{g^2 q^2}{2 \pi^2} N_* \left(1 + 
\mathcal{O}\left(\ln{N_*}/N_*\right)\right) \ .
\end{align}

In order to obtain the spectral index and other observables, we need to 
evaluate the slow-roll parameters
\begin{align}
 \epsilon = \frac{1}{2}\left( \frac{V'(\hat\sigma)}{V} \right)^2\ , \qquad 
 \eta = \frac{V''(\hat\sigma)}{V} \ , \qquad
\kappa =  -\frac{V'V'''(\hat\sigma)}{V^2} \ .
\end{align}
Here $\hat{\sigma}$ is the canonically normalized inflaton field which is
determined by (cf.~Eq.~(\ref{eq_lagrange}))
\begin{equation}
 \frac{d\sigma}{d \hat{\sigma}} = \frac{1}{\sqrt{K_{S \bar{S}}}} \ .
\end{equation}
On the inflationary trajectory
the derivatives of the scalar potential with respect to $\hat \sigma$  
can be written as ($n = 1,2,...$)
\begin{align}
\frac{d^n V}{d^n \hat \sigma^2} 
= \frac{d}{d \sigma}\left(\frac{dV^{n-1}}{d\hat\sigma}\right) 
\frac{d \sigma}{d \hat \sigma}    \ , 
\label{eq_dV}
\end{align}
from which one obtains the slow-roll parameters
%\begin{equation}
\begin{align}\label{eq_epsilon_eta}
\epsilon &\simeq \frac{1}{2} \left( \frac{g^2 q^2}{4\pi^2}\right)^2 \frac{1}{\sigma^2} 
\frac{1}{1 + \frac{1}{6} \chi (1 + \chi) \sigma^2} \ , \nonumber\\
\eta &\simeq - \frac{g^2 q^2}{4 \pi^2} \frac{1}{\sigma^2} 
\frac{(1 - \frac{1}{6} (1 + \chi) \sigma^2)(1 + \frac{1}{3} \chi (1 + \chi) 
\sigma^2)}{(1 + \frac{1}{6} \chi (1 + \chi) \sigma^2)^2} \ , \\
\kappa &\simeq -\left(\frac{g^2 q^2}{4\pi^2}\right)^2 \frac{2}{\sigma^4} 
\frac{(1 - \frac{1}{6}(1 + \chi)\sigma^2)(1 + \frac{1}{2}\chi (1 + \chi) 
\sigma^2(1 + \frac{2}{9} \chi (1 + \chi)\sigma^2(1 - \frac{1}{12}(1 + \chi) 
\sigma^2)))}{(1 + \frac{1}{6}\chi(1 + \chi)\sigma^2)^4} \ . \nonumber
\end{align}
%\label{eq_epsilon_eta}
%\end{equation}
In the large field regime, where
$\chi (1 + \chi) \sigma_*^2/6 > -(1 + \chi) \sigma_*^2/6 > 1$, the connection
between $\sigma_*$ and $N_*$ is given by Eq.~\eqref{largefield}, which
implies
\begin{equation}
\begin{split}
\epsilon_* & \simeq 3 \left( \frac{g^2 q^2}{4\pi^2}\right)^2 
\frac{1}{\chi^2\sigma_*^4} \simeq 
\frac{3}{4N_*^2} \ , \\
\eta_* &\simeq \frac{g^2 q^2}{2 \pi^2} \frac{1}{\chi\sigma_*^2} 
\simeq - \frac{1}{N_*} \ , \\
\kappa_* &\simeq  -4 \left( \frac{g^2 q^2}{4 \pi^2}\right)^2 
\frac{1}{\chi^2\sigma_*^4} \simeq 
 - \frac{1}{N_*^2} \ .
\end{split}
\label{eq_epsilon_eta}
\end{equation}
One easily verifies that these relations hold in the parameter range
\begin{align}
\frac{1}{-\chi} < \frac{gq}{2\sqrt{3}\pi} < 1\ .
\end{align}

Given these expressions, one then obtains for the scalar spectral index, the 
tensor-to-scalar ratio and the running of the spectral index ($N_* = 55$)
\begin{equation}
\begin{split}
n_s &\simeq 1 + 2\eta_* - 6\epsilon_* \simeq 1 - \frac{2}{N_*} \\
&\simeq  0.9636 \qquad [0.963 \pm 0.007] \ , \\
r &\simeq 12 \epsilon_* \simeq \frac{12}{N_*^2} \\
&\simeq 0.0040 \qquad [< 0.26] \ , \\
dn_s/d \ln k &\simeq -16\epsilon_*\eta_* + 24\epsilon_*^2 + 2\kappa_*
\simeq -\frac{2}{N_*^2} \\
&\simeq - 0.00066 \qquad [-0.022 \pm 0.010] \ .
 \end{split}
\label{eq_epsilon_eta}
\end{equation}
To leading order in $1/N_*$
the expressions agree with those of the Starobinsky model.
For comparison, the results obtained by the Planck collaboration 
\cite{Ade:2013uln,Ade:2013xla} are given in brackets. 
The agreement between predictions and observations is remarkable.
The amplitude of the scalar contribution to the primordial fluctuations
is given by
\begin{align}
A_s =\frac{1}{12\pi^2}\frac{V^{3}}{V'^2}\Big|_{\sigma =\sigma^*} 
 \simeq \frac{V_0}{18\pi^2}N_*^2 \ .
\end{align}
For $g^2 = 1/2$ the observed amplitude $A_s = (2.18 \pm 0.05)\times 10^{-9}$ 
\cite{Ade:2013xla} fixes the 
parameter $\xi$ to a value of order the GUT scale\footnote{At the end of 
hybrid inflation, cosmic strings are formed.
For these values of $g$ and $\xi$, and $q=8$, the string tension 
is $G\mu \simeq 3.16 \times 10^{-7}$, which
is marginally consistent with the recent Planck limit
$G\mu < 3.2 \times 10^{-7}$ \cite{Ade:2013xla} 
(see discussion in Ref.~\cite{Buchmuller:2012ex}).}, 
$\sqrt \xi \simeq 7.7\times 10^{15}\mathrm{GeV}$.
Note that the relative theoretical uncertainty of the slow-roll parameters is 
$\sim \ln N_*/N_* \sim 0.07$.

\vspace{0.6cm}
\noindent \textbf{Discussion}
\vspace{0.3cm}
\begin{figure}[t]
\center
\includegraphics[width = 0.8\textwidth]{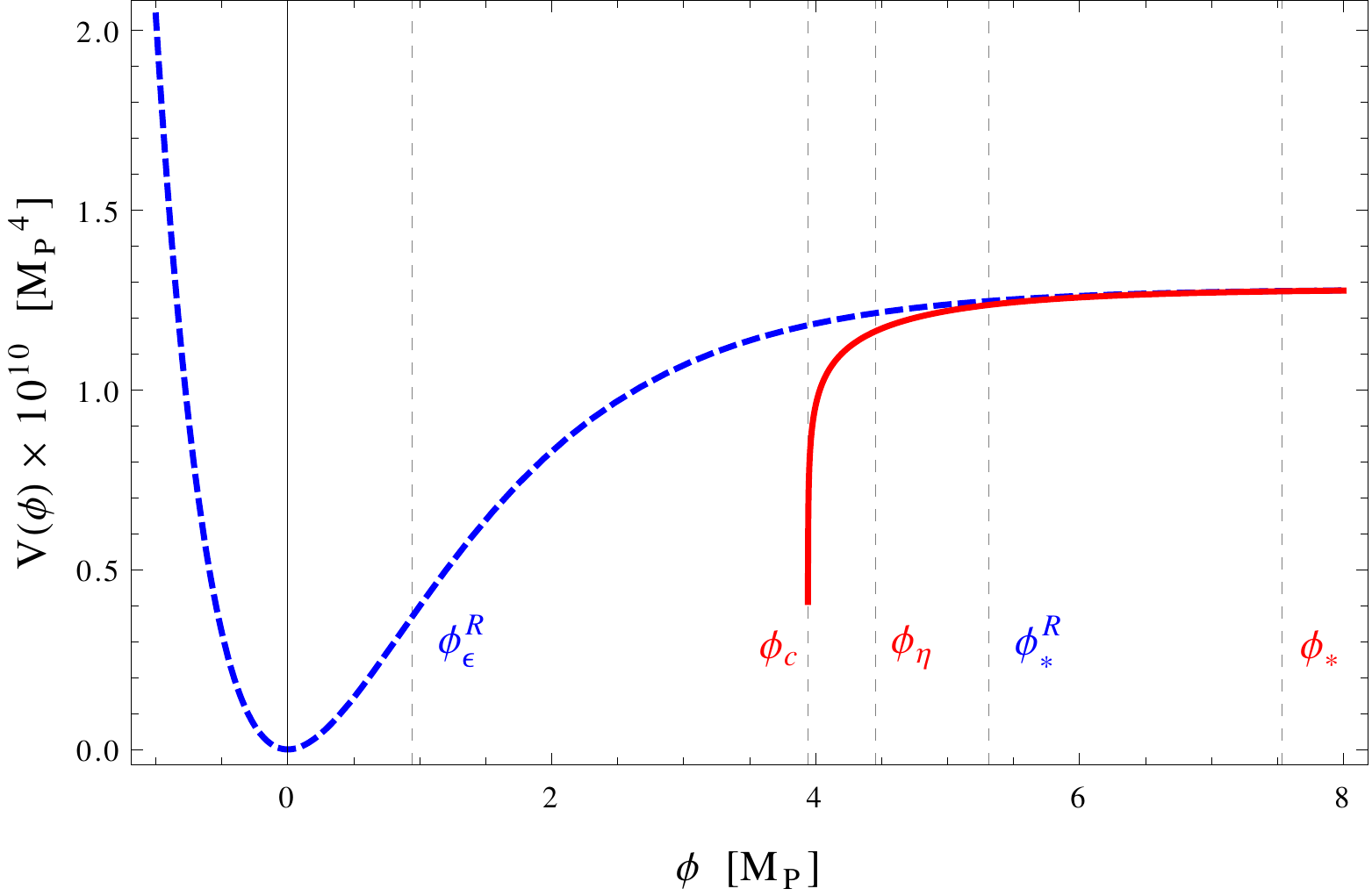}
\caption{Comparison of $R^2$ inflation (dashed line) and superconformal
D-term inflation 
(solid line) for $gq = 4\sqrt2$, $\lambda = 1$, $\chi = -10$ and $N_* = 55$. The
slow-roll regimes are $[\phi_{\epsilon}^R,\phi_*^R]$ and $[\phi_\eta,\phi_*]$,
respectively.}
\label{fig_comparison}
\end{figure}

Let us now discuss in more detail the connection between $R^2$ inflation and
superconformal hybrid inflation. The Starobinsky model 
\begin{align}
\frac{1}{\sqrt{-g}}\hat{\mathcal{L}}_R = 
\frac{1}{2} \left(R + \frac{1}{6M^2}R^2\right)
\end{align} 
is conveniently rewritten as scalar-tensor theory \cite{Whitt:1984pd},
\begin{align}
\frac{1}{\sqrt{-g}}\mathcal{L}_R \simeq 
\frac{1}{2} R 
- \frac{1}{2} g^{\mu\nu}
\partial_{\mu}\phi \partial_{\nu}\phi  - 
\frac{3}{4}M^2 \left(1-\exp{\left(-\sqrt{\frac{2}{3}}\phi\right)}\right)^2\ .
\end{align} 
For large values of $\phi$ one has
\begin{align}\label{VR}
V_R \simeq
\frac{3}{4}M^2 \left(1- 2\exp{\left(-\sqrt{\frac{2}{3}}\phi\right)}\right)\ .
\end{align}

In our discussion of D-term inflation we have used the field $\sigma$
which has a field dependent kinetic term. In the large field regime 
the connection with the canonically normalized field $\hat{\sigma}$
is given by
\begin{align}
\frac{d\sigma}{d\hat{\sigma}} = \frac{1}{\sqrt{K_{S\bar{S}}}}
\simeq \frac{\sqrt{6}(1-\frac{\chi}{6}\sigma^2)}{(-\chi\sigma)} \ ,
\end{align}
from which one obtains after a convenient choice of integration constant,
$\hat{\sigma}=\phi+\phi_0$,
\begin{align}
\sigma^2 = -\frac{6}{\chi}\left(C
\exp{\left(\sqrt{\frac{2}{3}}\phi\right)} - 1\right) \ .
\end{align}
Inserting this relation into the hybrid potential \eqref{hybridpotential}
one finds for large field values
\begin{align}
V \simeq
V_0 \left(1- 2\exp{\left(-\sqrt{\frac{2}{3}}\phi\right)}\right)\ ,
\end{align}
with
\begin{align}
V_0 = \frac{g^2}{2}\xi^2\left(1 +\mathcal{O}(g^2q^2\ln(g^2q\xi\chi))\right)\ ,
\quad C = \frac{g^4q^2\xi^2}{32\pi^2V_0}\ ,
\end{align}
which agrees with the potential \eqref{VR} after matching the constants,
$M^2 = 4 V_0/3$. 

The two potentials are compared in Figure~1. Although they almost coincide at
large $\phi$ corresponding to $N_* = 55$, they are completely different
at small $\phi$. Hence, also the two slow-roll
regimes, $[\phi_{\epsilon}^R,\phi_*^R]$ and $[\phi_{\eta},\phi_*]$,
differ significantly.
The slow-roll parameters agree up to higher orders in $1/N_*$, 
as discussed above, which corresponds to $\phi_*^{\rm R} \approx \phi_*$ and $V_R(\phi_*^{\rm R})
\approx V(\phi_*)$. 

Recently, the potential of $R^2$ inflation has also been derived from a
supergravity model with no-scale K\"ahler potential and Wess-Zumino
superpotential with specific couplings \cite{Ellis:2013xoa}. There are also
supergravity models with nonminimal couplings to gravity, which have the same 
behaviour as the Starobinsky model at large field values \cite{Linde:2011nh}. 
Another interesting example is Higgs inflation which, in the Einstein frame,
yields the scalar potential \eqref{VR} with 
$3M^2 = \lambda/{\hat{\chi}}^2 \equiv \hat{\lambda}$,
where $\hat{\chi}$ is the nonminimal coupling of the Higgs field to gravity
\cite{Bezrukov:2007ep}.

Why do all these models have the same asymptotic behaviour at large fields
in the Einstein frame? Consider Higgs or $R^2$ inflation in the Jordan frame.
After a field redefinition $\phi \rightarrow h(\phi)$ and a Weyl
transformation one obtains \cite{Bezrukov:2007ep}
\begin{align}\label{LagJ}
\frac{1}{\sqrt{-g_J}}\mathcal{L}_J^{\rm higgs} \simeq 
\frac{1}{2}(1 + \hat{\chi}h^2) R_J 
- \frac{1}{2} g_J^{\mu\nu}
\partial_{\mu}h \partial_{\nu}h  - \frac{\lambda}{4} h^4 \ . 
\end{align} 
Correspondingly, for superconformal D-term inflation one has 
\begin{align}\label{LagJ}
\frac{1}{\sqrt{-g_J}}\mathcal{L}_J^{\rm sc} &= 
\frac{1}{2}\Omega_0^{-2} R_J - \frac{1}{2} g_J^{\mu\nu}
\partial_{\mu}\sigma \partial_{\nu}\sigma  - \Omega_0^{-4}\ \frac{g^2}{2}\xi^2
\qquad \textrm{with} \nonumber\\
\Omega_0^{-2} &= 1 - \frac{1}{6}(1+\chi)\sigma^2 \ .
\end{align}
In the large field regime the two Lagrangians are identical after the
identification $h=\sigma$, $\hat{\chi} = -(1+\chi)/6$ and 
$\lambda = (1+\chi)^2g^2\xi^2/18$,
\begin{align}\label{LagJ}
\frac{1}{\sqrt{-g_J}}\mathcal{L}_J^{\rm sc} &\simeq 
-\frac{1}{12}(1+\chi)\sigma^2 R_J - \frac{1}{2} g_J^{\mu\nu}
\partial_{\mu}\sigma\partial_{\nu}\sigma  
- \frac{1}{72} (1+\chi)^2 g^2\xi^2 \sigma^4\ .
\end{align}
It is remarkable that in the large field regime in this Jordan frame, 
the Lagrangian is universal and scale invariant. The ratio of couplings,
$\hat{\lambda}=\lambda/{\hat{\chi}}^2 = 2 g^2\xi^2$ is fixed by observation to 
$2\times 10^{-11}$. For Higgs inflation, i.e. 
$\lambda = \mathcal{O}(1)$, this fixes the nonminimal coupling to
$\hat{\chi} \simeq 10^{5}$. In the case of superconformal D-term inflation
the dependence on the nonminimal coupling cancels, and for a GUT gauge
coupling, i.e. $g^2 = 1/2$, one obtains for $\xi$ the GUT scale,
$\sqrt{\xi} = 7.7\times 10^{15}\mathrm{GeV}$.  

It is surprizing that in the large field regime superconformal D-term 
inflation coincides with the Starobinsky model. As we showed, this is
a combined effect of quantum corrections and supergravity corrections
to scalar masses, which are determined by the superconformal K\"ahler
potential, and essentially independent of the size of the nonminal
coupling of the inflaton to gravity. It is remarkable that all models 
showing the asymptotic behaviour of the Starobinsky model are scale 
invariant at large field values in this Jordan frame, which appears 
to be the essence of $R^2$ inflation. 

\vspace{5mm}
\vspace{0.6cm}
\noindent \textbf{Acknowlegements}
\vspace{0.3cm}
%
%\noindent
%{\bf\large Acknowledements}\\

The authors thank J\'er\^ome Martin and Alexander Westphal for helpful
discussions.
This work has been supported by the German Science Foundation (DFG) within 
the Collaborative Research Center 676 ``Particles, Strings and the Early
Universe''.

%\newpage

%\markboth{}{}

\end{document}